\documentstyle[prl,aps,multicol,epsf]{revtex}

\begin{document}

\draft
%\widetext
\title{ Coulomb blockade of tunnelling through compressible rings formed
around an antidot: an explanation for $h/2e$ Aharonov-Bohm oscillations }

\author{M.~Kataoka,$^{1}$ C.~J.~B.~Ford,$^{1}$ G.~Faini,$^{2}$
D.~Mailly,$^{2}$ M.~Y.~Simmons,$^{1,\ast}$ and D.~A.~Ritchie$^{1}$}

\address{
$^{1}$Cavendish Laboratory, Madingley Road, Cambridge CB3 0HE, United Kingdom
}

\address{
$^{2}$Laboratoire de Microstructures et de Microelectronique - CNRS,
196,~Avenue Henri Ravera, 92220~Bagneux, France }

\date{\today}

\maketitle

\widetext

\begin{abstract}

\leftskip 54.8pt

\rightskip 54.8pt

We consider single-electron tunnelling through antidot states using a
Coulomb-blockade model, and give an explanation for $h/2e$ Aharonov-Bohm
oscillations, which are observed experimentally when the two spins of the
lowest Landau level form bound states. We show that the edge channels may
contain compressible regions, and using simple electrostatics, that the
resonance through the outer spin states should occur twice per $h/e$ period.
An antidot may be a powerful tool for investigating quantum Hall edge states
in general, and the interplay of spin and charging effects that occurs in
quantum dots.


\end{abstract}

\begin{multicols}{2}

\narrowtext

An antidot is usually considered as simply a smooth potential hill in a
two-dimensional electron gas (2DEG). In the quantum Hall regime, electrons
skip around its edge (in edge states), giving rise to single-particle (SP)
states with discrete energies. When the antidot is in the centre of a
constriction, the conductance through the constriction oscillates as a
function of perpendicular magnetic field $B$. This has been interpreted as
being due to periodic resonances through the SP energy levels. However, we
shall show that such an interpretation is too simplistic. It is useful to
compare the antidot with a closed semiconductor dot, in which Coulomb
blockade (CB) dominates the energy spectrum. The spectrum is modified by
confinement and interaction energies \cite{BEENAKKER}, and is further
complicated at high $B$ when edge states form in the dot. Compressible and
incompressible regions (CRs and IRs, respectively) \cite{BEENAKKER2,CHANG}
have been invoked to explain CB results in a dot \cite{McEUEN}, and
calculated for the boundaries of a 2DEG \cite{CHKLOVSKII}. In CRs, the
self-consistent potential is flat because of the screening of the external
electric field, and the highest occupied Landau level (LL) is pinned at the
Fermi energy $E_{\rm{F}}$. Nowadays, this has become a standard model when
2DEG edges are treated in a magnetic field. However, it has always been
assumed that such CRs should not exist around an antidot
\cite{FORD,MACE,SACHRAJDA,GOLDMAN,MAASILTA}.

The Aharonov-Bohm (AB) effect quantises the area of orbits around an antidot
so that each encloses an integer number of $h/e$ units of magnetic flux.
Because the potential is sloping, the energies of the states become discrete.
A small increase in $B$ shifts each SP state inwards to keep the enclosed
flux constant, sweeping each through $E_{\rm{F}}$ in turn with a period
$\Delta B = h/eS$, where $S$ is the area of the orbit at $E_{\rm{F}}$. When
$E_{\rm{F}}$ aligns with an SP state, electrons can tunnel resonantly between
nearby leads via the antidot (ignoring interactions). The resonance gives a
peak or dip in the conductance $G$ through the constrictions depending on the
direction of tunnelling \cite{MACE}. In this simple non-interacting picture,
there cannot be CRs because there would then be multiple SP states at
$E_{\rm{F}}$, and so $G$ would not oscillate since the antidot would always
be on resonance. However, this picture fails to explain certain phenomena
that have been observed, such as double-frequency ($h/2e$) AB oscillations
\cite{FORD,SACHRAJDA}. As $B$ increases, conductance peaks/dips often split
into pairs. This has naturally been interpreted as spin-splitting of AB
resonances, as it only happens when at least two spin-split LLs form closed
states around an antidot. As $B$ increases further, the splitting of
resonances saturates when the separation becomes exactly half the $h/e$
period, showing $h/2e$ periodicity. The amplitudes of each oscillation in a
pair also become equal, although the different tunnelling distance for each
spin should result in different amplitudes. It has been suggested that this
should be the result of the charging of antidot states \cite{FORD,SACHRAJDA}.
However, it was an unresolved issue whether charging should really occur in
such a system \cite{FORD,GOLDMAN}. 

Recently, we reported that an antidot shows net charge oscillations as a
function of $B$, and tunnelling through the bound states is Coulomb blockaded
\cite{KATAOKA}. However, even if the charging energy $E_{\rm{C}}$ is taken
into account, the non-zero Zeeman splitting and the SP energy spacing $\Delta
E_{\rm{sp}}$ prevent the oscillations of two spins from becoming exactly out
of phase, unless $E_{\rm{C}}$ greatly exceeds the other energies. In our
previous experiments, it was estimated that $E_{\rm{C}}$ saturates at around
150~$\mu$eV at high $B$, which is only an order of magnitude larger than
$\Delta E_{\rm{sp}}$ there \cite{KATAOKA}.

In this paper, we explain the $h/2e$ oscillations with a CB model. Firstly,
experimental evidence is presented to show that the resonances involve
states of only one spin. Then, a self-consistent potential with CRs and IRs
is considered for the system, and antidot charging is explained in terms of
simple electrostatics. It is found that the screening in each CR causes a
resonance through the outer spin states with $h/2e$ periodicity. Imperfect
screening at lower $B$ may lead to a pairing of resonances through the outer
spin states.

% FIG. 1.

\begin{figure}[t]

\epsfxsize=85truemm

\centerline{\epsffile{ 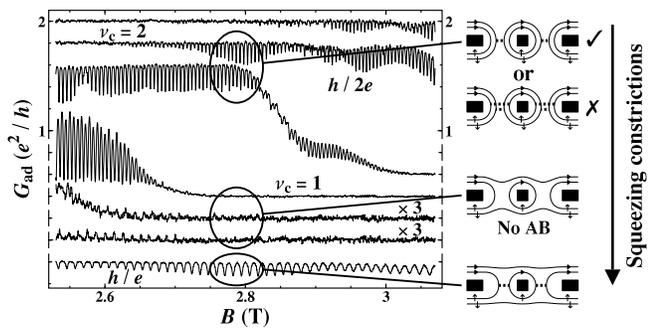}}

\vspace{6pt}

\caption{AB conductance oscillations: the two constrictions were squeezed
symmetrically between traces, which are offset by $0.2e^{2}/h$ down the page
for clarity. Two curves are expanded by a factor of 3, as indicated. The
diagrams at right show the geometry of edge channels (solid lines) for
$B \approx 2.8$~T. Black boxes indicate surface gates. Tunnelling between
edge channels is represented by a dotted line. }

\label{fig:DBHIGH}

\end{figure}

The details of the antidot device and experiment are presented elsewhere
\cite{KATAOKA}. Figure~\ref{fig:DBHIGH} presents the antidot conductance
$G_{\rm{ad}}$ measured in the $h/2e$ oscillation regime (top curves) by
gradually squeezing the constrictions (by increasing the side-gate voltages
negatively) whilst keeping the symmetry, for a constant antidot voltage (the
curves are offset vertically). In the top curve around 2.8~T, the filling
factor in each constriction was $\nu_{\rm{c}}=2$, while in the lowest curve,
$\nu_{\rm{c}}<1$ ($G_{\rm{ad}} = \nu_{\rm{c}} e^2/h$). As the constrictions
become narrower, the $h/2e$ oscillations vanish and $G_{\rm{ad}}$ shows the
$\nu_{\rm{c}}=1$ plateau, then eventually $h/e$ oscillations appear. The
latter are obviously due to resonant reflection through states of the lower
spin, say spin up ($\uparrow $) (see bottom diagram at right). If the
resonance through these inner states contributes to the $h/2e$ oscillations,
there is no apparent reason why it stops between the $h/2e$ and $h/e$
oscillations where the conductance shows the $\nu_{\rm{c}}=1$ plateau. Thus,
we conclude that also in the $h/2e$ regime, there should be no resonance
through the inner states, so {\em all} the resonances must be through the
outer states [spin down ($\downarrow $)]. At this stage, the inner states
must be too far from the current-carrying edges for electrons to tunnel. 
Similar behaviour has also been observed for $B$ as low as 1.7~T, where
irregular $h/2e$ oscillations are present.

In order to understand the mechanism of the $h/2e$ oscillations, it is
necessary to treat the electrostatics of the antidot edge in detail. In the
conventional non-interacting picture, the antidot potential increases
smoothly towards the centre, and the SP states are filled up to $E_{\rm{F}}$.
The carrier density as a function of the distance from the antidot centre is
therefore step-like: an abrupt change occurs where a LL crosses $E_{\rm{F}}$.
Thus the width of the change should be of order the magnetic length $l_{B} =
\sqrt{\hbar/eB}$. Such a carrier distribution is unfavourable
electrostatically, especially at large $B$, where the deviation is large. In
bulk edges, CRs form, causing flat plateaux in the self-consistent potential.
As mentioned earlier, in antidot systems, it has always been considered that
the potential cannot be completely flat at $E_{\rm{F}}$. However, if there is
CB, so that single-electron tunnelling into and out of the antidot edge only
occurs when the net charge can jump between $\pm e/2$, then a conductance
oscillation will still be possible \cite{EP2DS}, just as in a metallic dot
where the SP energy is negligible.

% FIG. 2.

\begin{figure}

\epsfxsize=80truemm

\centerline{\epsffile{ 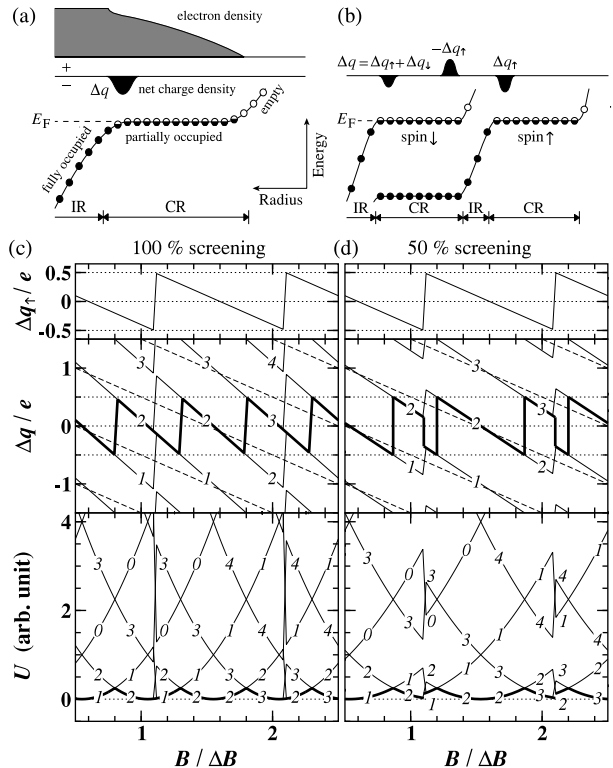}}

\vspace{6pt}

\caption{(a) Schematic diagram of a LL around an antidot with a CR
surrounded by an IR. (b) When two CRs exist of opposite spins, the outer CR
tries to screen the charging of the inner spin($\uparrow$), therefore,
$-\Delta q_{\uparrow}$ piles up at the inner edge of the outer ring. As a
result, the net charge $\Delta q$ built up at the outer edge of the CR is the
sum of the charging of both spins. Charge oscillations and the capacitive
energy $U$ are shown for (c) 100 \% screening and (d) 50 \% screening, with
an arbitrary origin of $B$.}

\label{fig:ChModel}

\end{figure}

Let us consider the case in which only one LL encircles the antidot. The
self-consistent LL would be as shown in Fig.~\ref{fig:ChModel}(a). Here,
unlike in bulk edges, electrons in the compressible states are trapped around
the antidot and surrounded by the $\nu=1$ IR, which contains one electron per
state, and so the total number of electrons in the CR must be an integer
(assuming no hybridisation with the CR on the other side of the IR). The IR
acts as a potential barrier, through which an electron must tunnel from a
lead to reach the antidot edge. As $B$ increases, all the antidot states move
towards the centre. The electrons in the CR are forced to occupy a smaller
area, pushed by the surrounding IR. This results in a shift of the electrons
from the initial distribution, leading to a build-up of negative charge.
Because of the compressibility and the ring shape, the net charge $\Delta q$
appears only at the outer edge of the CR as shown in the figure. This is
similar to the fact that a net charge on a metal sphere or cylinder appears
only on the outer surface.

$\Delta q$ cannot relax until it reaches $-e/2$, when a single electron can
escape from the CR making $\Delta q = +e/2$. This is the origin of the CB of
tunnelling through the antidot states. This phenomenon very much resembles CB
in a metallic dot. Because many partially occupied states are at
$E_{\rm{F}}$, the SP energy is not important. One should note that after the
resonance, the number of electrons in the CR remains the same despite one
escaping, because another electron is supplied to its outer edge from the
surrounding IR by shifting the CR back to its position just after the
previous resonance. The period of the oscillation should be determined by the
area enclosed by the outer edge of the CR.

We now consider what happens when two spins form antidot states. As shown in
Fig.~\ref{fig:ChModel}(b), this results in the formation of two compressible
rings separated by the narrow $\nu=1$ IR, if the spin-splitting is
sufficient. These will act rather like the parallel plates of a capacitor.
Here, the rings are treated as infinite, concentric cylinders because the IR
should be narrower than the thickness of the 2DEG ($\sim 100-200$ \AA ). The
net charge of the inner spin state $\Delta q_{\uparrow }$ will induce an
image charge $- \Delta q_{\uparrow }$ at the inner edge of the outer CR, as
shown, to keep the potential flat in the outer CR. Thus, another net charge
which is the sum of the net charges in the two CRs, $\Delta q = \Delta
q_{\uparrow } + \Delta q_{\downarrow }$, appears at the outer edge of the
outer CR (Gauss' theorem). Here, $\Delta q_{\downarrow }$ is the net charge
in the spin $\downarrow$ CR, which includes the contribution from fully
occupied spin $\uparrow$ states there due to a slight increase in LL
degeneracy [its dependence on $B$ is shown as dashed lines in the middle
panel, Fig.~\ref{fig:ChModel}(c)]. As $B$ increases, $\Delta q_{\uparrow }$
performs saw-tooth oscillations as for the case of a single CR, regardless of
the outer spin states [top panel, Fig.~\ref{fig:ChModel}(c)]. As a result,
the net charge $\Delta q$ in the outer edge of the spin $\downarrow$ CR
changes as shown by solid lines in the middle panel. Its capacitive energy $U
\propto (\Delta q)^2$ is plotted in the bottom panel, each number indicating
a particular occupation pattern, shifted by one state from the previous
pattern. When the inner spin state is on resonance and one electron escapes,
$\Delta q$ becomes more positive by $e$, and so the outer CR must acquire one
electron to keep $\Delta q$ below $+ e/2$. This is effectively a spin-flip
process leading to an oscillation of $\Delta q$ accompanied by two resonances
per $h/e$ period as shown by the bold lines. Here, the $h/e$ period is
determined by the outer edge of the outer CR. This is our explanation for the
pure $h/2e$ oscillations. Because the resonances are all via the outer CR,
the amplitudes should be the same.

\begin{figure}

\epsfxsize=70truemm

\centerline{\epsffile{ 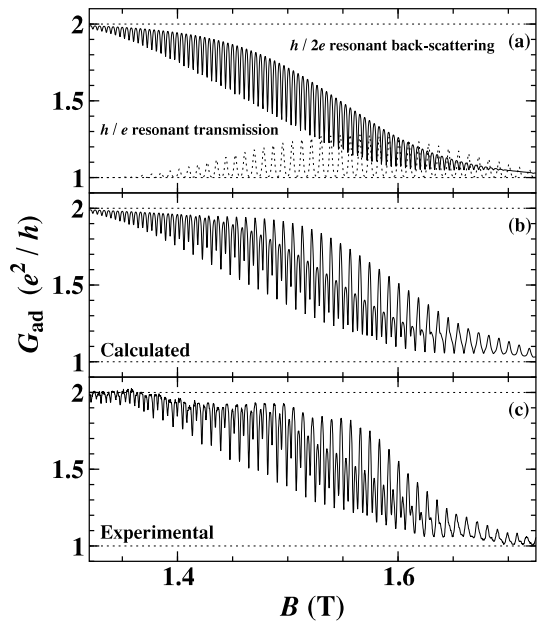}}

\vspace{6pt}

\caption{Modelling irregular $h/2e$ AB oscillations experimentally observed
between $\nu_{\rm{c}}=2$ and 1 plateaux (the lowest curve). See text for
detail.}

\label{fig:DbModel}

\end{figure}

\begin{figure}

\epsfxsize=75truemm

\centerline{\epsffile{ 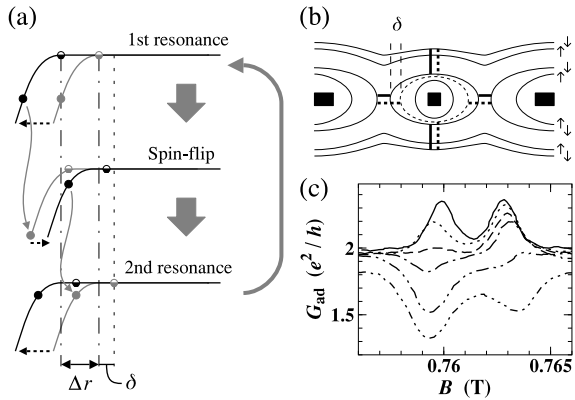}}

\vspace{6pt}

\caption{(a) The movement of the LL of the outer spin states as $B$
increases. Only the outer part is shown. (b) More realistic picture of edge
channels around an antidot. Since the potential slope is shallower in the
constrictions, the edge channels are further apart inside them. The positions
of the outer edge channel at the 1st and 2nd resonances are shown by solid
and dashed lines, respectively. (c) $G_{\rm{ad}}$ measured as both
constrictions are squeezed from $\nu_{\rm{c}}=2$ by increasing the
antidot-gate voltage $V_{\rm{g}}$. The curves are offset horizontally to
compensate for the shift of the resonance positions with $V_{\rm{g}}$. }

\label{fig:LowBModel}

\end{figure}

The new model can be extended to lower $B$, where conductance oscillations
show complicated features. Figure~\ref{fig:DbModel}(c) presents irregular
$h/2e$ AB oscillations observed between the $\nu_{\rm{c}}=2$ and 1 plateaux
at moderate $B$. These were originally interpreted as alternate
back-scattering resonances of two spins, from the presence of apparent pairs
of dips. However, this cannot explain why the conductance oscillations of the
inner spin vanish when the conductance goes down to the $\nu_{\rm{c}}=1$
plateau. In our new model the resonant reflection (RR) process through the
outer spin states should give $h/2e$ oscillations [solid line,
Fig.~\ref{fig:DbModel}(a), generated from a series of differentiated Fermi
functions]. Also, there should be resonant transmission (RT) due to inter-LL
scattering from the next LL (which is not transmitted) via the inner LL
[vertical lines in Fig.~\ref{fig:LowBModel}(b)]. RT through the outer spin
states will just offset the RR, but that through the inner spin states will
have period $h/e$ [dotted line, Fig.~\ref{fig:DbModel}(a)]. This can be seen
as peaks coming off the $\nu_{\rm{c}} = 1$ plateau at the high $B$ end of the
experimental data. The period of the RT should not be exactly twice that of
the resonance through the inner spin because of the difference in radii.
Adjusting the period difference and the phase, and adding the two curves, we
obtain Fig.~\ref{fig:DbModel}(b), with a striking resemblance to the
experimental data. For the best fit, the periodicity was increased linearly
across the range by 7\% and 5\% for the inner and outer spin states,
respectively.

% FIG. 3.

Despite the good agreement, the curves are qualitatively different in some
respects. The large peaks are mainly due to RT through the inner CR, and are
sharp in the calculation, whereas they are rounded near the $\nu_{\rm{c}}=2$
plateau in the experimental data. Also, it is not obvious why the resonances
die out as the conductance reaches the plateau. This is the case at whatever
field the plateau is reached. This may be explained if tunnelling into the
inner CR is prevented when the outer CR is completely isolated (even at the
constrictions), i.e. $\nu_{\rm{c}}=2$. This might be due to dynamic screening
as the electron tunnels past that CR. The dips at the low $B$ end seem to be
paired, which could be as a result of only partial screening: if the outer CR
cannot perfectly screen the charging of the inner CR, the exact $h/2e$
periodicity is broken as shown in Fig.~\ref{fig:ChModel}(d) for 50\%
screening ($\Delta q = 0.5 \Delta q_{\uparrow } + \Delta q_{\downarrow }$).
There are still two resonances per $h/e$ period, but they are paired. This
requires the relative phase of the charge oscillations of the inner and outer
states to be such that a spin-flip occurs between the pairs, otherwise, there
will be only one resonance in the period. 

% FIG. 4.

As $B$ decreases, the spin-split CRs should merge, and it is not clear when
our model will break down. Even when the CRs have merged, the spin $\uparrow$
states will be incompressible in the outer part of the CR. Any excess charge
with this spin will therefore accumulate further in, whereas any charging of
the spin $\downarrow$ states will occur at the outer edge. Thus the outer
spin $\downarrow$ states may still partially screen the charging of the spin
$\uparrow$ states. At low $B$, the non-interacting picture describes the
antidot behaviour well \cite{MACE}. Figure~\ref{fig:LowBModel}(c) presents
the measured antidot conductance at various dot-gate voltages, with
$\nu_{\rm{c}}$ decreasing from 2. The left peak becomes a dip as intra-LL
scattering, across the constrictions, takes over from inter-LL scattering,
which occurs at the edges facing the bulk \cite{MACE}. The right peak behaves
similarly, when the constrictions are narrower. The pairs of peaks/dips were
previously interpreted as spin-split resonances, i.e.\/ alternate resonances
through inner and outer spin states \cite{MACE}. This requires that the
tunnelling distances for two adjacent back-scattering resonances be
different. In our model, paired resonances through the outer CR per $h/e$
period can readily be explained by imperfect screening. However, the origin
of the different amplitudes for the two resonances is more subtle.  The
spatial positions of the antidot states are slightly different for the first
and second resonances (defined so that a spin-flip occurs in between) as
shown schematically in Fig.~\ref{fig:LowBModel}(a). The difference in
positions is $\delta =\Delta r \Delta B_{\rm{s}}/\Delta B$, where $\Delta r$
is the radius difference between neighbouring states and $\Delta B_{\rm{s}}$
is the spacing between the resonances in $B$. Because $\Delta r$ decreases as
$1/B$, $\delta$ becomes negligible at large $B$, giving rise to pure $h/2e$
oscillations. Also, since the actual antidot potential is steeper at the
edges facing the bulk 2DEG than in the constrictions [see
Fig.~\ref{fig:LowBModel}(b)], the inter-LL scattering may hardly be affected.
In contrast, for the back-scattering process, the amplitude at the first
resonance will be larger than at the second one. Note that the original
explanation \cite{MACE} requires a very similar difference in tunnelling
distance.

Although it may be possible to describe the curves in our model, it is
unclear whether CRs should really exist at such low $B$. There may be a
gradual transition from a non-interacting picture to the new picture, as $B$
increases. If CRs do exist, it may be necessary to reconsider the analysis of
the excitation spectra (around 1.4~T) in our previous paper \cite{KATAOKA},
or else, the CRs may not yet have formed at that field. It may be too
simplistic to consider tunnelling into individual SP states -- the details of
the many-body states ought to be considered. Also, the potential shown here
is just schematic -- since the charge that forms on either side of an IR
around the antidot is usually much less than $e/2$, antidot charging up to
$\pm e/2$ will cause the potential to vary strongly within each period.

In summary, we have proposed a mechanism of antidot charging which gives rise
to the pure $h/2e$ AB oscillations which we have observed experimentally.
When both spin states of the lowest Landau level encircle the antidot, they
form two concentric compressible regions.  Screening in those regions, and
Coulomb blockade, then force the resonances through the outer compressible
region to occur twice per $h/e$ cycle. Irregular $h/2e$ oscillations observed
at lower fields can also be explained using the same picture. Our results and
interpretation in this paper show that the period and lineshapes of the AB
oscillations seen in transport provide detailed information about the various
spin states and the Coulomb interaction between them. An antidot may
therefore be a powerful tool for investigating edge states in the quantum
Hall regime, and the interplay of spin-polarisation and charging effects.
Such information cannot be obtained reliably in quantum dots, but, since
there should be many similarities between the effects in dots and antidots,
it may, for example, be possible to investigate the physics of maximum
density droplets \cite{MDD}, or of the spin textures seen in dots
\cite{SACHRAJDAHAWRYLAK}. The period-doubling of AB oscillations observed in
quantum dots \cite{DBdot} may be explained by our model. Even in extended
edges, the way in which one edge state screens its neighbour may determine
how current flows there.

This work was funded by the U.~K. EPSRC. We thank C.~H.~W.~Barnes,
C.~W.~J.~Beenakker, I.~Larkin and J.~R.~Yates for useful discussions. M.~K.
acknowledges financial support from the Cambridge Overseas Trust.

$^{\ast}$Present address: School of Physics, University of New South Wales,
Australia.

\end{multicols}

\end{document}